\documentclass[12pt]{iopart}

\usepackage{aas_macros}
\usepackage{graphicx}
\usepackage{color}
\usepackage{natbib}
\usepackage{iopams}  
\usepackage{url}
\usepackage{textcomp}

\begin{document}

\title[Pixel-based spectral characterization of mid-IR arrays ]{Pixel-based spectral characterization of mid-infrared Si array detectors for astronomical observations in space}

\author{Takuro Tsuchikawa$^1$, Hidehiro Kaneda$^1$, Daisuke Ishihara$^2$, Takuma Kokusho$^1$, Takehiko Wada$^2$}

\address{$^1$ Graduate School of Science, Nagoya University, Furo-cho, Chikusa-ku, Nagoya, Aichi 464-8602, Japan}
\address{$^2$ Institute of Space and Astronautical Science, Japan Aerospace Exploration Agency, 3-1-1 Yoshinodai, Sagamihara, Kanagawa 252-5210, Japan}
\ead{tsuchikawa@u.phys.nagoya-u.ac.jp}
\vspace{10pt}
\begin{indented}
\item[March 2020]
\end{indented}

\begin{abstract}
Mid-infrared (IR) array detectors have been used for astronomical observations in space. However, the uniformities of their spectral response curves have not been investigated in detail, the understanding of which is important for spectroscopic observations using large array formats. We characterize the spectral responses of all the pixels in IR array detectors using a Fourier transform infrared spectrometer and cryogenic optics for measurements at high signal-to-noise ratios. We measured the spectral responses of  the Si:As impurity band conduction (IBC) array, a flight back-up detector for $\it AKARI$/IRC. As a result, we find that the Si:As array has intrinsic variations in the spectral response along the row and column directions of the array. We also find that the cutoff wavelength of the Si:As IBC array depends on the intensity of the incident light. 
\end{abstract}

 \vspace{2pc}
\maketitle
%
%
\renewcommand{\thefootnote}{\arabic{footnote}}

\section{Introduction}
Mid-infrared (IR) array detectors have been used in space for focal-plane instruments aboard IR astronomical satellites, such as $\it ISO$/ISOCAM \citep{Cesarsky1996}, $\it {Spitzer}$/IRAC \citep{Fazio2004}, MIPS \citep{Rieke2004}, IRS \citep{Houck2004}, $\it {AKARI}$/IRC \citep{Onaka2007} and $\it {WISE}$ \citep{Mainzer2008}. They are also used for future IR space telescopes such as $\it {JWST}$/MIRI \citep{Rieke2015} and $\it {SPICA}$/SMI \citep{Kaneda2018}. 
The first mid-IR array detector was developed more than three decades ago, which had small array formats (e.g., 32 $\times$ 32 Si:Bi by \citealt{Arens1983} and 16 $\times$ 16 Si:Bi by \citealt{Lamb1984}). Now large format ones (e.g., 1k ${\times}$ 1k Si:As IBC for $\it {WISE}$) are available, which enable us to carry out observations with high efficiency.

The mid-IR array detectors of various instruments have been evaluated for the uniformities of the photometric responses which were integrated over the wavelengths of their IR filter bands \citep[e.g.,][]{Ressler2008}.
\citet{Ressler2008} reported that the uniformity of the photometric response in the array varies from pixel to pixel by $3\%$ for the Si:As array detector, and found a fringe pattern of the photometric response in the array, which was likely to be caused by swirl dislocations in the crystalline structure of the silicon boule.
On the other hand, variations of the spectral response curves from pixel to pixel have not been investigated in detail, the understanding of which is important for spectroscopic observations using large array formats. 
In this paper, we characterize the spectral responses of all the pixels in IR array detectors.

\section{Method\label{method}}
\subsection{Mid-IR array detector}

For the purpose of characterizing the pixel-based spectral responses, we evaluate a Si:As impurity band conduction (IBC) array, a flight back-up detector manufactured by Raytheon Vision Systems, Santa Barbara, USA \citep{Estrada1998}, for the mid-IR channels of InfraRed Camera \citep[IRC;][]{Onaka2007} aboard $\it {AKARI}$ \citep{Murakami2007}. 
The Si:As array has a format of 256 $\times$ 256, the characteristics of which are summarized in table~\ref{tab:det}. 
The pixel size is ${30\times30\,{\rm {\mu}m}}^2$, and the array size is ${7.68\times7.68\,{\rm mm}}^2$. 
An anti-reflection (AR) coating is applied to the detector surface to increase the quantum efficiency.
The frame readout rate is designed to be low ($11\;{\rm Hz} $) because of the need to reduce the heat dissipation. 
The array has a hybrid structure consisting of an IR detector semiconductor and silicon-based cryo-CMOS readout integrated circuit (ROIC), CRC-744, which consists of source follower per detector (SFD) input circuits. We integrate charge carriers with a detector capacitance and perform non-destructive readouts. The effective voltage bias, $\rm {\it V}_{bias}$, of 0.6 V is applied to the detector when the reset switch is turned on, which is the difference between the voltage levels of the unit cell SFD gate bias, $\rm {\it V}_{rstuc}$ that is the reset level, and the detector common bias connected to the detector backside, $\rm {\it V}_{det}$.

\begin{table}
\caption{Characteristics of the Si:As IBC/CRC-744.}             
\label{tab:det}
\begin{indented}
\item[]\begin{tabular}{@{}ll}
\br
   Parameter & Si:As IBC/CRC-744\\ 
\mr                      
  Wavelength range &  $5-27\;{\rm {\mu}m}$ \\ 
  Format &  $256 \times 256$ \\ 
  Pixel size & $30\;{\rm {\mu}m}\times30\;{\rm {\mu}m}$ \\ 
  Well capacity$^{\rm a}$  & $>1\times10^5\;{\rm e^-}$ \\ 
  Operating temperature &  $<\,10\;{\rm K}$ \\
Number of outputs & 4 \\ 
  Quantum Efficiency (QE) & $>50\;\%$ \\   
  Frame rate & $11\;{\rm Hz} $ \\
\br
\end{tabular}
\item[] $^{\rm a}$ At $\sim$ 0.6 V for the voltage bias applied to the detector.
\end{indented}
\end{table}
  
\subsection{Experimental configuration}
We evaluate the pixel-based spectral responses of the Si:As array using a Fourier transform infrared spectrometer (FT-IR), Bruker VERTEX 70v, which is efficient to evaluate array detectors since the spectra of the incident light from the FT-IR to all the pixels in the array can be measured simultaneously with signal-to-noise ratios ($\it S/N$) higher than those from dispersive spectrometers such as monochrometers. 
The FT-IR is composed of an IR continuum source and a $\rm RockSolid^{TM}$ interferometer, consisting of dual cube corner mirrors which are least affected by mirror tilts. 
A globar lamp and a Mylar filter of $6\;{\rm {\mu}m}$ in thickness are used for the light source and the beam splitter, respectively, which are selected considering the wavelength coverage of the Si:As array.

\begin{figure}
		\begin{center}
		\includegraphics[width=13cm,clip]{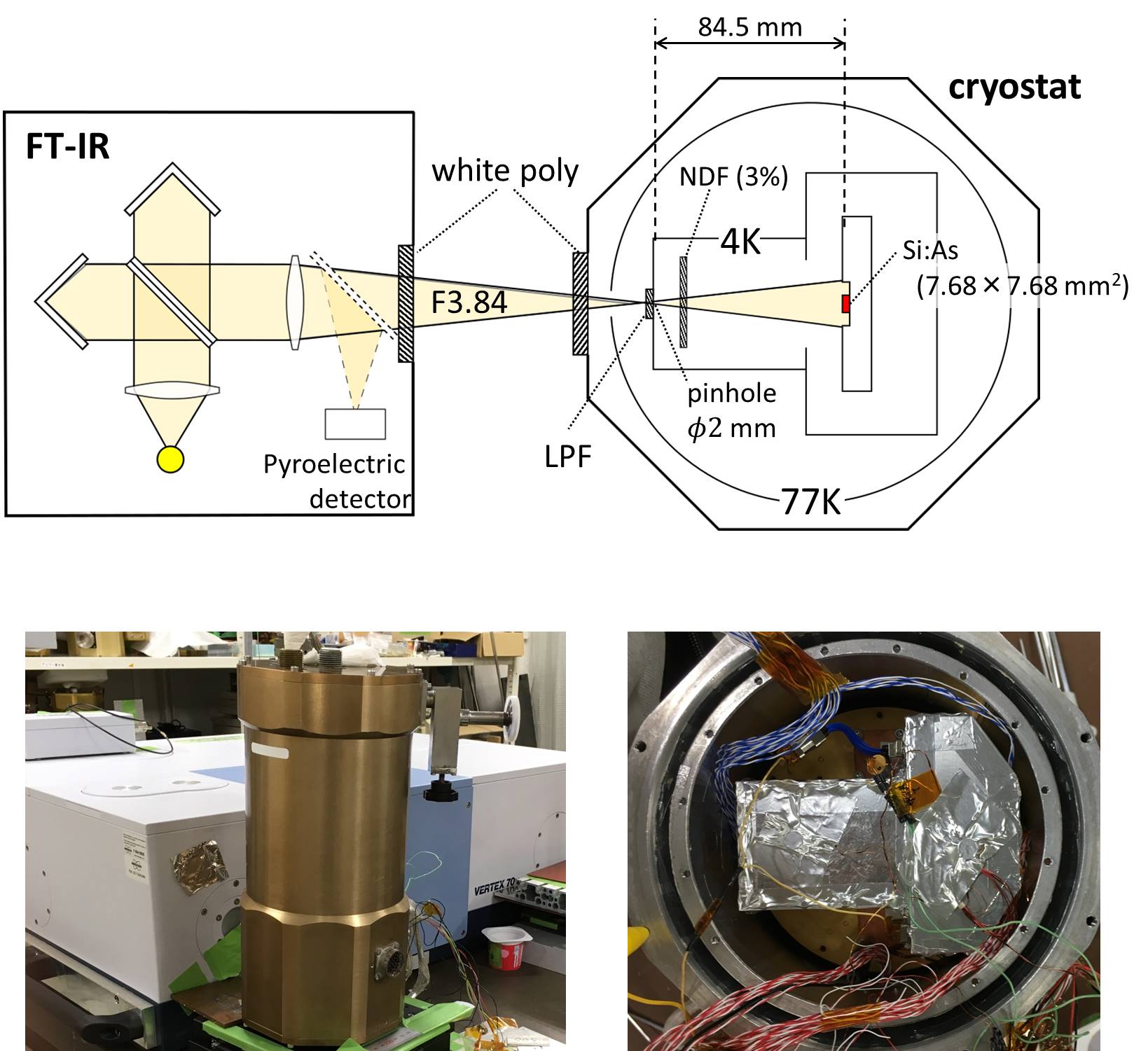}
		\caption{$\it {Top\;panel}$: schematic view of the configuration of the measurement for the Si:As array. The light path shown with the dashed line is used for the measurement of the source spectrum. $\it {Bottom\;left\;panel}$: a photo of the setup for the measurement. The light from the FT-IR is guided into the cryostat placed in front of the FT-IR.  $\it {Bottom\;right\;panel}$: the setup for the measurement inside the cryostat. The Si:As array is installed within the 4~K shield.} 
              \label{fig:sias_config}
        \end{center}
    \end{figure}

\begin{figure}
		\begin{center}
		\includegraphics[width=10cm,clip]{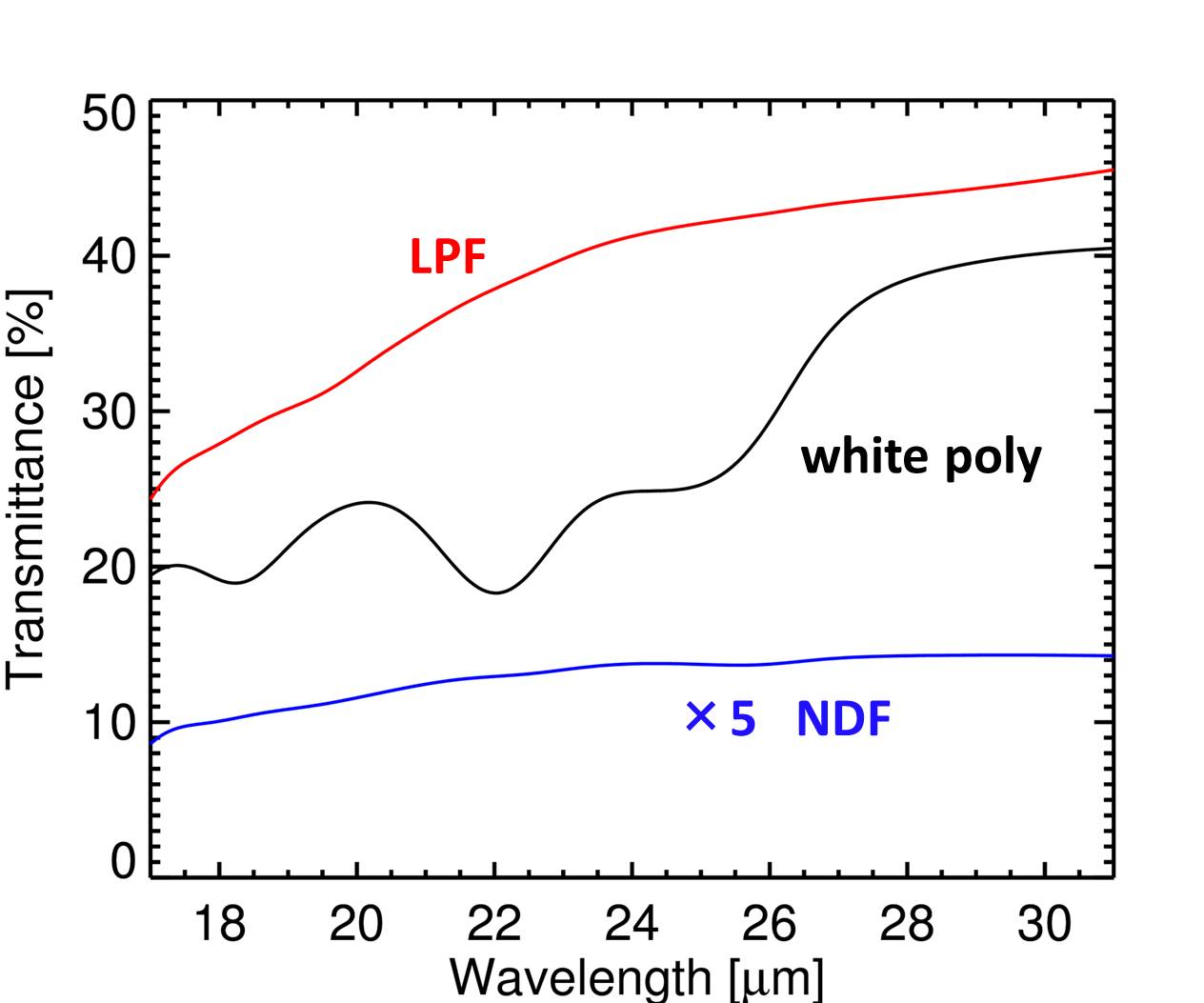}
		\caption{Transmission curves of the windows and the filters used in the present measurement. Red, black and blue lines show the transmission curves of the long wavelength-pass filter, the white polyethylene windows and the neutral-density filter, respectively. The transmission curve of the neutral-density filter is scaled upward by a factor 5.} 
              \label{fig:transmit}
        \end{center}
    \end{figure}

The Si:As array has to be operated at a cryogenic temperature (see table~\ref{tab:det}), and thus we cooled it using a LHe cryostat. 
Figure~\ref{fig:sias_config} shows our set-up for the measurements. 
The light from the FT-IR at a room temperature is to be guided into the cryostat at 4~K through white polyethylene windows, the transmission curve of which is shown in figure~\ref{fig:transmit}. %
Accordingly, the measurements are necessarily performed under relatively high background (BG) environments, where $300\;{\rm K}$ BG photon noise dominates. 
In addition, the Si:As array is saturated immediately due to the $\rm 300\;K$ BG radiation, since the IR detectors designed for space  applications for low BG have well capacities considerably smaller than those for ground-based ones (see table~\ref{tab:det}). 
Hence we cannot increase the total amount of the incident light to improve the $\it S/N$, which has to be optimized by using a neutral-density filter (NDF; figure~\ref{fig:transmit}) on the basis of the well capacity.

As shown in figure~\ref{fig:sias_config}, we adopt a configuration like a pinhole camera, the aperture size of which is set to be $2\;{\rm mm}$ in diameter. 
The BG light irradiates the detector only within the same solid angle as the light from the FT-IR, and therefore the configuration enables the best $\it S/N$ measurement under the same F-number (F3.84) of the light from the FT-IR. 
Moreover, the Mylar beam splitter does not transmit IR light but only for a broad wavelength range of 15--300$\;{\rm {\mu}m}$, and thus we use a long wavelength-pass filter (LPF; figure~\ref{fig:transmit}) whose cutoff wavelength is approximately 15$\;{\rm {\mu}m}$ in the cryostat to suppress only the BG light. 
For the purpose of reducing the stray light, the inner wall of the $4\;{\rm K}$ shield is coated with an IR low-reflectance paint (Nextel Black Velvet manufactured by Mankiewicz), and also labyrinth structures are adopted at the junctions of the components constituting the $4\;{\rm K}$ shield.

Separately from the response of the IR array, we also measure the spectrum of the source of the FT-IR with the pyroelectric detector 
located inside the FT-IR, whose spectral response is known to be 
flat. As shown in figure 1, before measuring the IR array, we 
inserted a mirror to the beam to redirect it to the pyroelectric 
detector. We obtain the reference spectrum, multiplying the source 
spectrum and the transmission curves of the windows and the filters. 
We confirm that the level of the repeatability of the reference 
spectrum obtained by the FT-IR measurements is smaller than $0.05\%$, 
at the wavelength of 20~$\mu$m, which is negligible compared to the 
measurement error of $>0.2\%$ (see below). We divide the spectra of 
the Si:As array by the reference spectrum to evaluate the spectral 
response, $R(\lambda)$. 

\begin{figure}
		\begin{center}
		\includegraphics[width=10cm,clip]{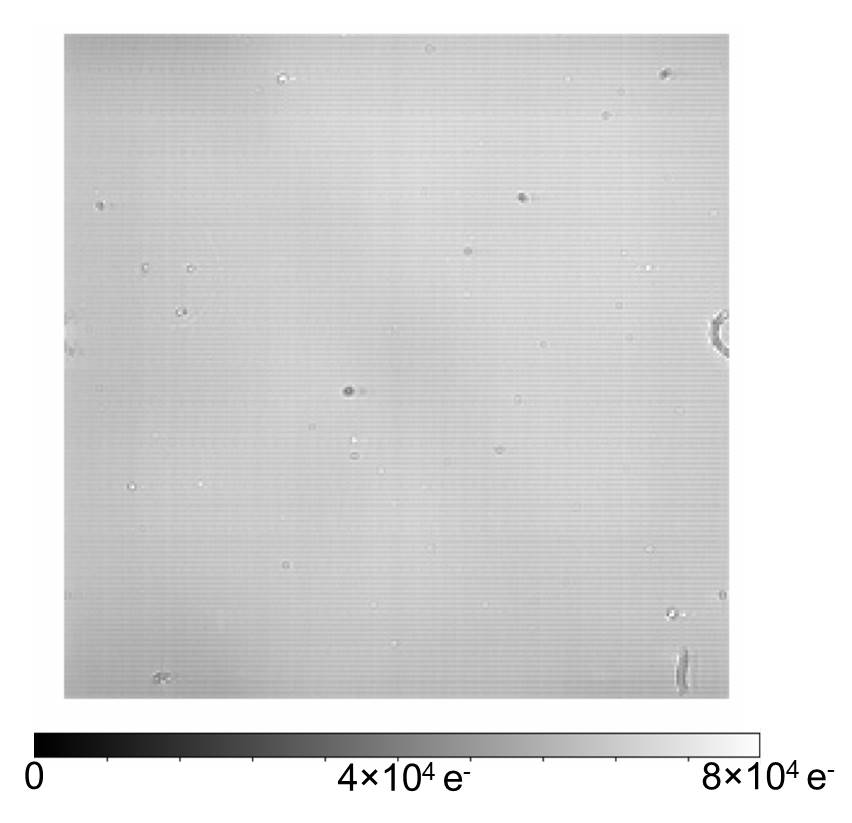}
		\caption{Intensity map of the incident light falling onto the Si:As array. The color bar is given in units of the number of electrons.}
              \label{fig:intensity_map}
        \end{center}
    \end{figure}

Figure~\ref{fig:intensity_map} shows the intensity map of the incident light falling onto the Si:As array under the above configuration, which shows that the incident light illuminates all the pixels almost uniformly.
It is noted, however, that the intensity of the incident light from the FT-IR is slightly weakened near the corners of the array, presumably due to imperfections of the optics of the measurement system including the FT-IR, where the spectrum of the incident light  is likely to be affected by the optical diffraction effect. 
Therefore we repeated the measurement of the spectral responses under 6 configurations with different illuminated positions and combined them by intensity-weighted average in order to eliminate the contributions from those likely problematic. Furthermore, by spatial high-pass filtering, we focus only on the components of the spectral responses with spatial frequencies higher than the low spatial frequency typical of the intensity variation shown in figure~\ref{fig:intensity_map}, which corresponds to the spatial scale of 11 pixels.

\subsection{Detector operations and data processing\label{sec:data_process}}
We read out the integration signals of all the pixels of the Si:As array repeatedly, while moving the scan mirror at a constant speed, ${\it v}_{\rm m}$. 
It should be noted, however, that the sampling rate of the Si:As array, $11\,{\rm Hz}$, is significantly lower than the Nyquist frequency of the interferograms, $137\,{\rm Hz}$, estimated from both ${\it v}_{\rm m}{\sim}0.1\,{\rm cm/s}$ and the shorter cutoff wavelength, 15$\,\rm{\mu}m$, and hence under-sampled interferograms are expected to be obtained, which cause the spectral aliasing.
Thus we adopt 1/16 frame readout, which was used for the scan mode operation of the $\it {AKARI}$/IRC all-sky survey \citep{Ishihara2006}, instead of the standard frame readout, and performed the measurement 16 times to cover the whole frame.
The sampling rate of the 1/16 frame readout is $148~{\rm Hz}$, and thus we can obtain Nyquist-sampled interferograms.

The interferograms are expected to be somewhat smoothed from the real interferograms, since the optical path difference (OPD) of the interferometer changes while charge carriers are integrated by the Si:As array. 
Below we neglect the non-linearity of signal integration; we have confirmed that the non-linearity level is as low as $<10^{-3}$ for the present data, which is estimated from the full-range integration curve data measured separately. The corrections of the interferograms are conducted by considering the differences between the obtained and real interferograms as a result of convolution with a rectangular function, $U(t)$, the width of which corresponds to the integration time.
Hence the relation between the observed interferogram, $I(t)$, and the real interferogram, $I_{\rm real}(t)$, is given by $I(t) = I_{\rm real}(t)~{\ast}~U(t)$.

According to the convolution theorem, the spectrum, $F(k)$, obtained by Fourier-transforming the observed interferogram is the multiplication of the real spectrum, $F_{\rm real}(k)$, and the sinc function, ${\rm sinc}(k)$, which are the Fourier transforms of $I_{\rm real}(t)$ and the rectangular function, $U(t)$, respectively, as follows:
\begin{equation}
F(k) = F_{\rm real}(k){\times}{\rm sinc}(k),
\label{eq:spec}
\end{equation}
Therefore we can obtain $F_{\rm real}(\lambda)$ by dividing $F(\lambda)$ by the sinc function.
Finally, we estimate the measurement errors by calculating standard deviations for 60 measurements.

\section{Results}

\begin{figure}
		\begin{center}
		\includegraphics[width=10cm,clip]{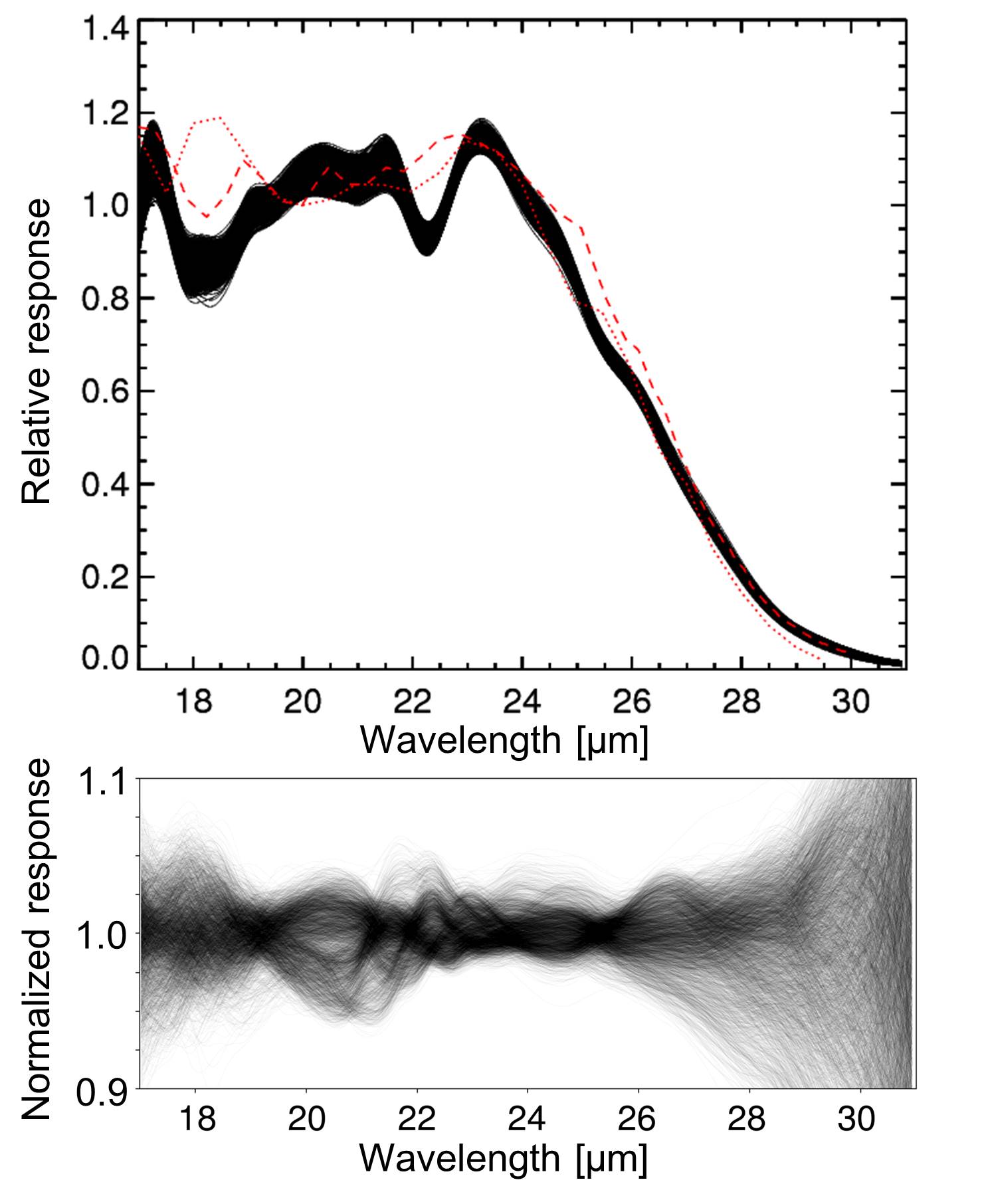}
		\caption{Spectral responses obtained for all the 256 $\times$ 256 pixels of the Si:As array. The upper and lower panels show the spectral responses normalized to the intensity of the light integrated over the wavelength range of 17 to 33 $\mu$m and those normalized to the median response at each wavelength, respectively. The red dotted and dashed lines in the upper panel represent the spectral responses normalized at 20~$\mu$m obtained by \citet{Lum1993} and \citet{Estrada1998}, respectively.}
              \label{fig:spec_map}
        \end{center}
    \end{figure}

Analyzing the interferograms measured with the method described in section~\ref{method} and combining all the results acquired with the different illuminated positions on the array, we obtain the spectral responses of all the pixels in the array.
The upper panel of figure~\ref{fig:spec_map} shows all the obtained spectral responses normalized with light intensities incident on each pixel, which are calculated by integrating the spectrum of each pixel over the wavelength range of 17 to 33 $\mu$m. As can be seen in this figure, the cutoff wavelengths are located around 27 $\mu$m, which is consistent with the spectral range listed in table~\ref{tab:det}. We compare them with the spectral responses obtained in the previous measurements by \citet{Lum1993} and \citet{Estrada1998}, which indicates that our measurement shows a fairly good agreement with the previous measurements. Furthermore, we find that the spectral responses are notably similar from pixel to pixel, yet varying significantly compared to the pixel-averaged measurement error of about 0.003 at $20\,{\rm {\mu}m}$ (see the lower panel of figure~\ref{fig:spec_map}).

\begin{figure}
		\begin{center}
		\includegraphics[width=14cm,clip]{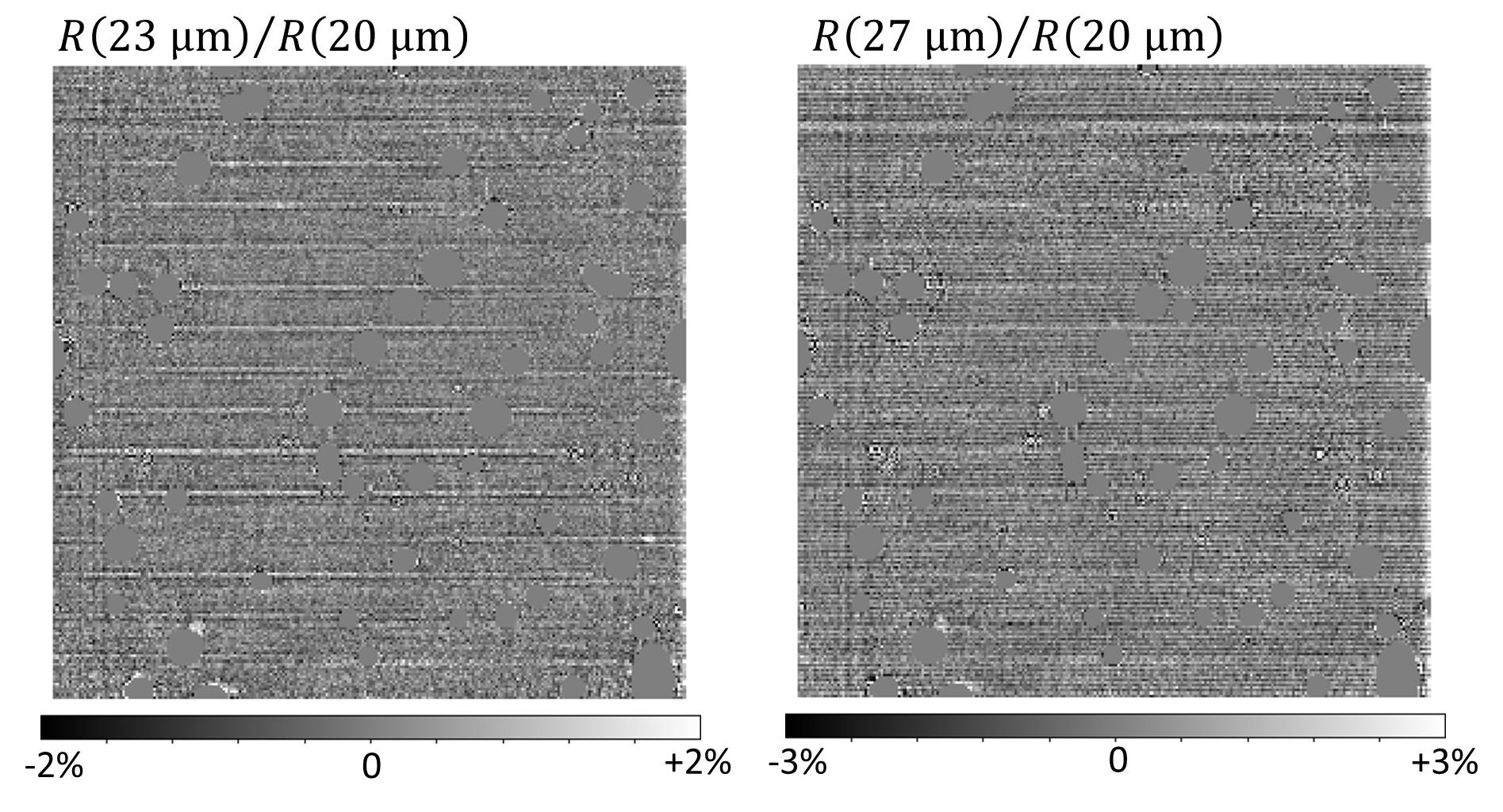}
		\caption{Spectral response ratio maps of $R(23~{\rm {\mu}m})/R(20~{\rm {\mu}m})$ and $R(27~{\rm {\mu}m})/R(20~{\rm {\mu}m})$ after removal of low spatial frequency components. The gray-scale bars are given in the fractional differences between the spectral response ratios and their averages in the array. We mask the regions likely to be affected by particulates. }
              \label{fig:HPF_map}
        \end{center}
    \end{figure}

In order to characterize the pixel-by-pixel variations of the spectral responses, we calculate two spectral response ratios, $R(23\,{\rm {\mu}m})/R(20\,{\rm {\mu}m})$ and $R(27\,{\rm {\mu}m})/R(20\,{\rm {\mu}m})$, for all the pixels in the array. The former represents the global gradients of the spectral responses, while the latter the cutoff wavelengths.
Figure~\ref{fig:HPF_map} shows the spectral response ratio maps of $R(23\,{\rm {\mu}m})/R(20\,{\rm {\mu}m})$ and $R(27\,{\rm {\mu}m})/R(20\,{\rm {\mu}m})$. We find that both maps have periodic spatial variations along the column direction.

\section{Discussions}

\subsection{Pixel-by-pixel variations of the spectral responses}
\begin{figure}
		\begin{center}
		\includegraphics[width=14cm,clip]{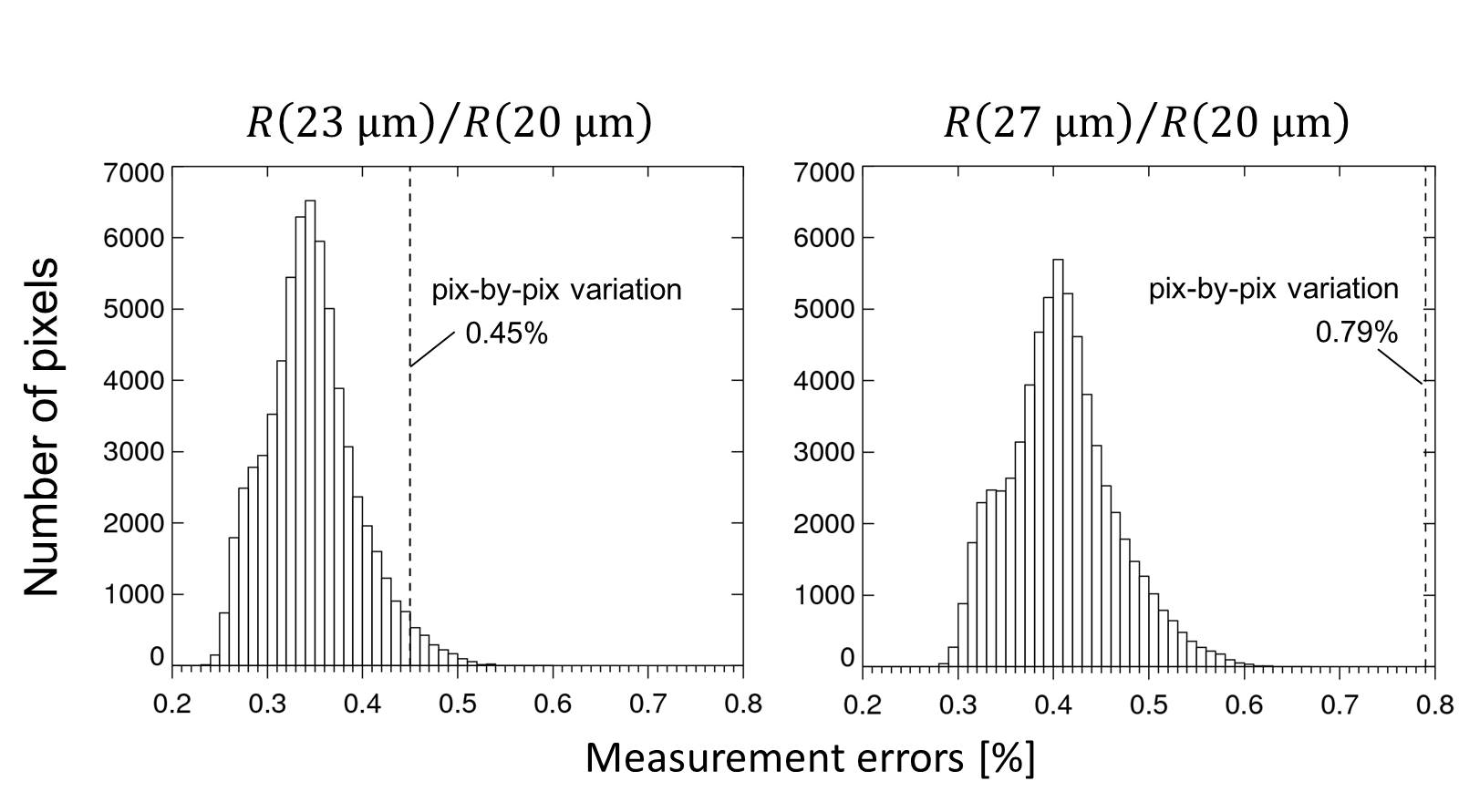}
		\caption{Histograms of the measurement errors for the $R(23\,{\rm {\mu}m})/R(20\,{\rm {\mu}m})$ and $R(27\,{\rm {\mu}m})/R(20\,{\rm {\mu}m})$ maps. Dashed lines correspond to the pixel-by-pixel variations of the maps in figure~\ref{fig:HPF_map}, which are calculated by the standard deviations of the ratios.}
              \label{fig:err_histo}
        \end{center}
    \end{figure}

We calculate the standard deviations of the spectral response ratios from the maps in figure~\ref{fig:HPF_map}. The resultant pixel-by-pixel variations are $0.45\%$ and $0.79\%$ for $R(23\,{\rm {\mu}m})/R(20\,{\rm {\mu}m})$ and $R(27\,{\rm {\mu}m})/R(20\,{\rm {\mu}m})$, respectively. 
We compare the pixel-by-pixel variations of the spectral response ratios with the measurement error for each pixel. Figure~\ref{fig:err_histo} shows the histograms of the measurement errors for the $R(23\,{\rm {\mu}m})/R(20\,{\rm {\mu}m})$ and $R(27\,{\rm {\mu}m})/R(20\,{\rm {\mu}m})$ maps, the pixel averages of which are $0.35\%$ and $0.41\%$, respectively. 
The pixel-by-pixel variations are shown as dashed lines in figure~\ref{fig:err_histo}, from which we find that only $3\%$ and $0\%$ of the pixels in the array show the measurement errors larger than the pixel-by-pixel variations of the $R(23\,{\rm {\mu}m})/R(20\,{\rm {\mu}m})$ and $R(27\,{\rm {\mu}m})/R(20\,{\rm {\mu}m})$ maps, respectively. Hence we find that the pixel-by-pixel variations of the spectral responses thus measured are intrinsic for almost all the pixels with sufficient $\it S/N$.

\begin{figure}
		\begin{center}
		\includegraphics[width=12cm,clip]{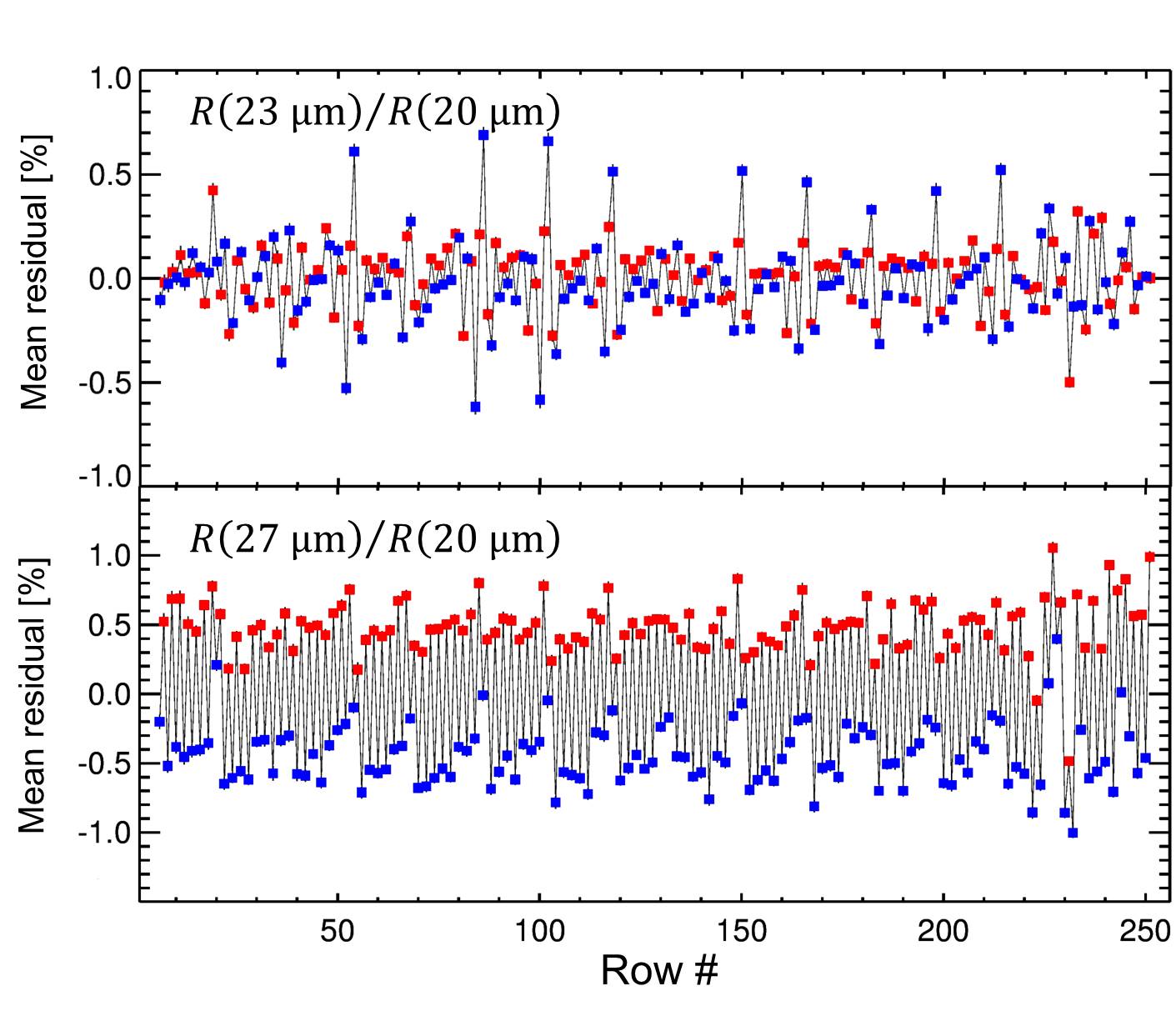}
		\caption{Distributions of the residuals of $R(23\,{\rm {\mu}m})/R(20\,{\rm {\mu}m})$ and $R(27\,{\rm {\mu}m})/R(20\,{\rm {\mu}m})$ averaged per row relative to those averaged for all the pixels. Red and blue squares represent the data points of the odd and even rows, respectively. }
              \label{fig:row}
        \end{center}
    \end{figure}

As can be seen in figure~\ref{fig:HPF_map}, the spectral response ratio maps of $R(23\,{\rm {\mu}m})/R(20\,{\rm {\mu}m})$ and $R(27\,{\rm {\mu}m})/R(20\,{\rm {\mu}m})$ show periodic spatial variations along the column direction.
To visualize the row-by-row and column-by-column variations more clearly, we average the maps for every row and column. 
The upper and lower panels of figure~\ref{fig:row} show the distributions of $R(23\,{\rm {\mu}m})/R(20\,{\rm {\mu}m})$ and $R(27\,{\rm {\mu}m})/R(20\,{\rm {\mu}m})$, respectively, averaged per row. 
It is found from the upper panel that $R(23\,{\rm {\mu}m})/R(20\,{\rm {\mu}m})$ shows appreciable variations every 16 rows. As described in section~\ref{sec:data_process}, we operated the 1/16 frame readout, in which 16 adjacent rows were read out simultaneously. Hence this operation is likely to cause the periodic spatial variation, presumably due to a slight change in the detector temperature immediately after the readout.

\begin{figure}
		\begin{center}
		\includegraphics[width=12cm,clip]{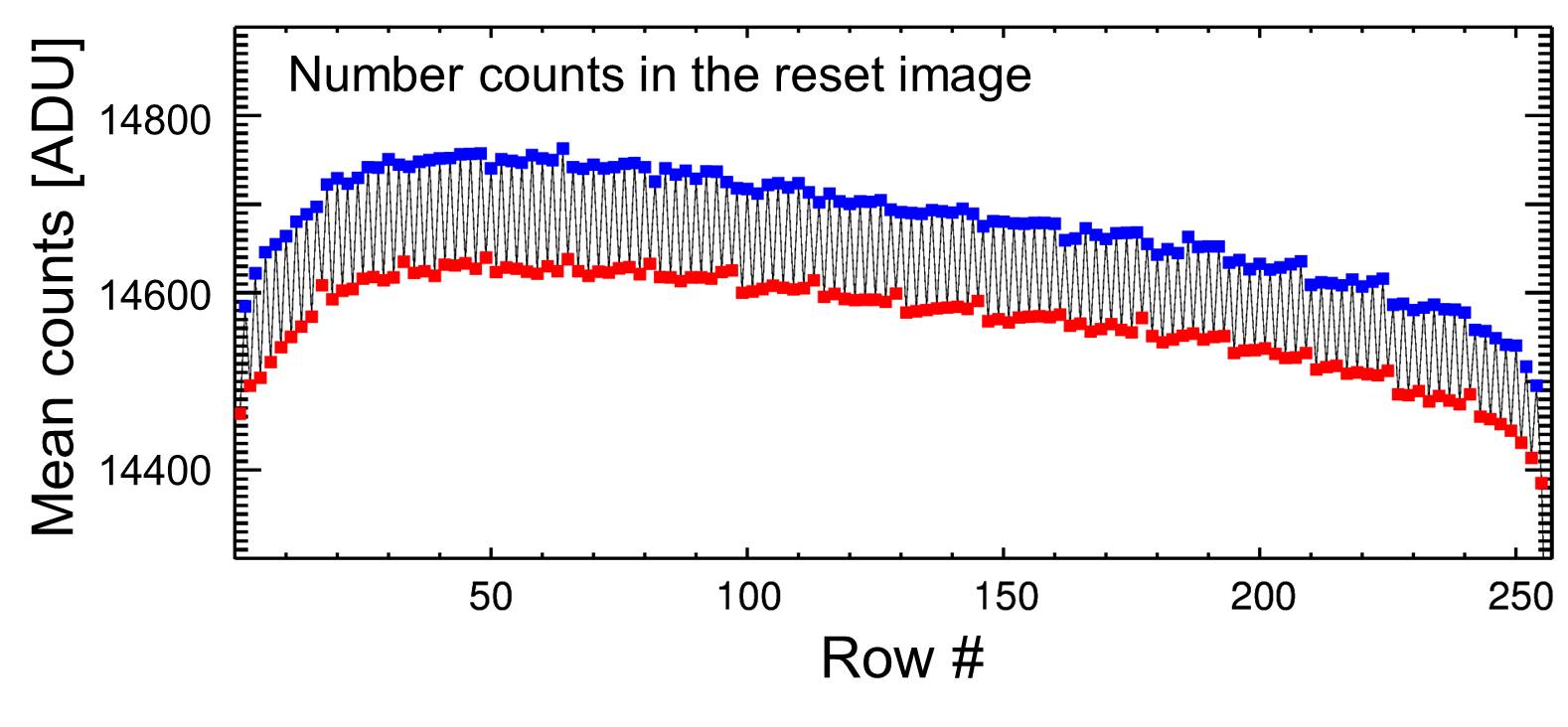}
		\caption{Distributions of the number counts in the reset image averaged per row. Red and blue squares represent the data points of the odd and even rows, respectively. }
              \label{fig:reset}
        \end{center}
    \end{figure}

The lower panel in figure~\ref{fig:row} clearly indicates that the spectral responses in the odd rows show systematically higher $R(27\,{\rm {\mu}m})/R(20\,{\rm {\mu}m})$ and thus longer cutoff wavelengths than those in the even rows. In general, spectral responses at longer wavelengths of the IBC-type detectors are known to increase with $\rm {\it V}_{bias}$, as mentioned in $\it Spitzer$/IRS Instrument Handbook Version 5.0. This is explained physically by the Poole-Frenkel effect, which lowers the effective Coulombic potential by the external field. 
In light of this, we speculate that there is a systematic difference in $\rm {\it V}_{bias}$ between the even and odd rows. In order to verify the hypothesis, we obtain an image with the reset switch turned on, to measure the spatial variation of $\rm {\it V}_{bias}$ in the array directly. As shown in figure~\ref{fig:reset}, we find that the odd rows in the reset image show values systematically lower than the even rows, which indicates that the unit cell SFD gate bias, $\rm {\it V}_{rstuc}$, is systematically higher in the odd rows than in the even rows. Hence the variations every 2 rows of the spectral responses in the lower panel of figure~\ref{fig:row} are likely to be caused by the Poole-Frenkel effect, considering $\rm {\it V}_{bias}={\it V}_{rstuc}- {\it V}_{det}$ with $\rm {\it V}_{det}$ constant.

\begin{figure}
		\begin{center}
		\includegraphics[width=12cm,clip]{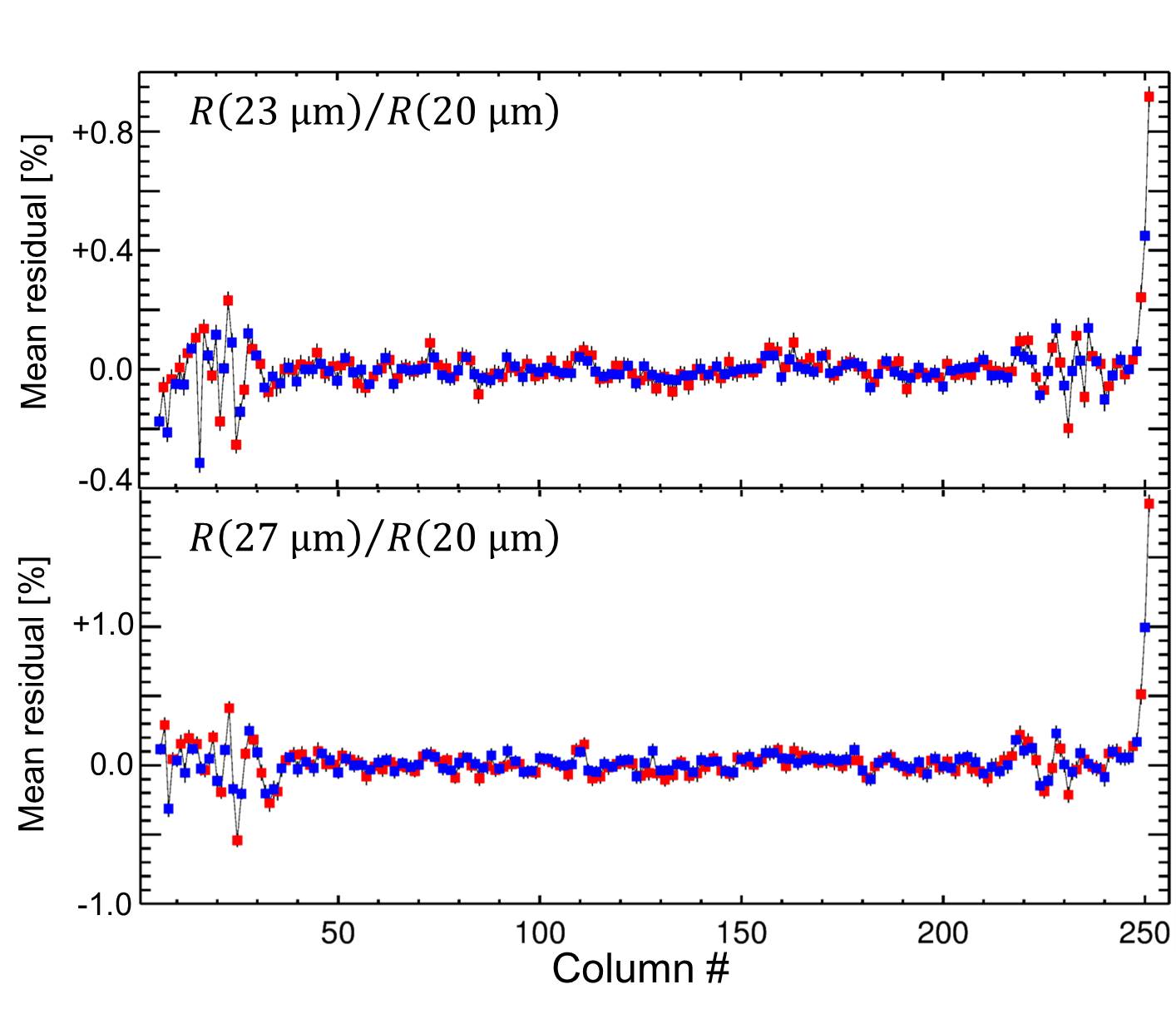}
		\caption{Distributions of the residuals of $R(23\,{\rm {\mu}m})/R(20\,{\rm {\mu}m})$ and $R(27\,{\rm {\mu}m})/R(20\,{\rm {\mu}m})$ averaged per column relative to those averaged for all the pixels. Red and blue squares represent the data points of the odd and even columns, respectively. }
              \label{fig:col}
        \end{center}
    \end{figure}

Figure~\ref{fig:col} shows the distributions of $R(23\,{\rm {\mu}m})/R(20\,{\rm {\mu}m})$ and $R(27\,{\rm {\mu}m})/R(20\,{\rm {\mu}m})$ averaged per column, which does not exhibit periodic spectral variations such as figure~\ref{fig:row}, except for wave-like variations near both ends of the columns. 
We find that the wave-like variations are spatially correlated between the distributions of $R(23\,{\rm {\mu}m})/R(20\,{\rm {\mu}m})$ and $R(27\,{\rm {\mu}m})/R(20\,{\rm {\mu}m})$, which indicate that the global gradients of the spectral responses are variable near both ends of the columns.
The variations of the global gradients of the spectral responses can be explained by the spatial variations of the thickness of the detector layer and/or the AR coating.
Another possibility is the ``picture frame effect" seen in some arrays (e.g., Figure~2-3 in the SpeX observing manual\footnote{See \url{http://irtfweb.ifa.hawaii.edu/~spex/spex_manual/SpeX_manual_21oct14.pdf}}). The picture frame noise is likely to be affected by detector temperature fluctuations at a milli-Kelvin level \citep{Bechter2019}. Thus the wave-like variations of the spectral responses can be explained by a slight difference in the detector temperature, similarly to the variations every 16 rows in the upper panel of figure~\ref{fig:row}.

\subsection{Dependency on the intensity of the incident light}
\begin{figure}
		\begin{center}
		\includegraphics[width=12cm,clip]{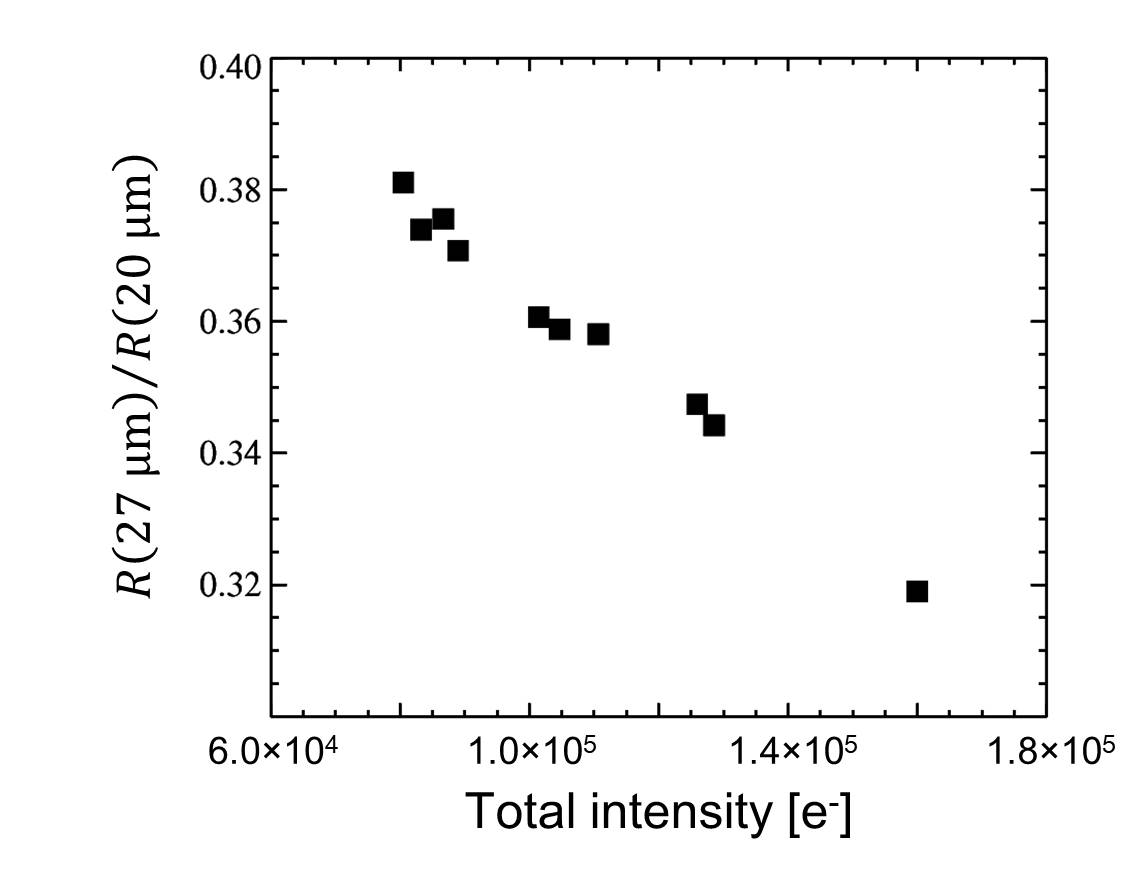}
		\caption{Averaged spectral response ratio $R(27\,{\rm {\mu}m})/R(20\,{\rm {\mu}m})$ as a function of the intensity of the incident light.}
              \label{fig:bias}
        \end{center}
    \end{figure}

We also measure the spectral responses under various BG levels by increasing the broad-band IR BG level. In this measurement, the 16 rows ($Y$=161--176) are read out to calculate the average spectral response ratio $R(27\,{\rm {\mu}m})/R(20\,{\rm {\mu}m})$.
In figure~\ref{fig:bias}, we show the average $R(27\,{\rm {\mu}m})/R(20\,{\rm {\mu}m})$ as a function of the pixel-averaged intensity of the incident light falling onto the Si:As array, which indicates the clear trend that the spectral responses have shorter cutoff wavelengths under higher BG environments. 
For the source follower per detector circuit adopted in mid-IR array detectors such as the Si:As IBC array, $\rm {\it V}_{bias}$ decreases with integrating the charge carriers. Therefore, during the charge integration, the spectral responses are expected to be changed by the Poole-Frenkel effect as mentioned in the above discussion. Thus the obtained spectral responses, which are regarded as the average spectral responses during the time of the charge integration, are expected to have shorter cutoff wavelengths under higher BG environments. This result demonstrates that, in order to obtain highly accurate spectra, we have to calibrate the spectral responses taking into account the light intensity for each pixel as well as the pixel-by-pixel intrinsic variations.

\section{Conclusion}

Understanding of the spectral responses of large-format infrared array detectors is important for spectroscopic observations in space astronomy. 
We have successfully characterized the pixel-by-pixel intrinsic variations of the spectral responses of the infrared array detector using the cryogenic optics and the FT-IR which enable the measurements at high $\it S/N$. We have found that the pixel-by-pixel variations of the spectral responses have systematic spatial variations along both row and column directions.
For the row-by-row variations, the spectral responses in the odd rows show longer cutoff wavelengths than those in the even rows, which is likely to be caused by the Poole-Frenkel effect due to the spatial variation of $\rm {\it V}_{bias}$.
For the column-by-column variations, global gradients of the spectral responses are variable near both ends of the columns, which can be explained by the spatial variations of the thickness of the detector layer.
Furthermore, we have found that spectral responses of IBC-type array detectors significantly change with the light intensity in the BG environments most probably due to changes in $\rm {\it V}_{bias}$ by the Poole-Frenkel effect during the charge integration. 
For future space observations using large-format IR arrays, we suggest the importance of evaluating pixel-based $\rm {\it V}_{bias}$ through the measurement of a reset image and taking into account it as well as source intensities for calibration of the spectral properties.

This work was supported by JSPS KAKENHI Grand Number JP19K03927, and is based on the backup detector for $AKARI$, a JAXA project with the participation of ESA.

\bibliographystyle{aasjournal}
\bibliography{SiAs_PASP}

\end{document}